\documentclass[aps,prd,onecolumn,showpacs,nofootinbib,superscriptaddress]{revtex4}  
\pdfoutput=1
\usepackage{graphicx}
\usepackage{epstopdf}
\usepackage{amsmath}
\usepackage{amsfonts}
\usepackage{amssymb}
\usepackage{appendix}
\usepackage{comment}
\usepackage{bbold}
\usepackage{slashed}
\usepackage{subfigure}
\usepackage{setspace}
\usepackage{footnote}
\usepackage{multirow}
\usepackage{longtable}
\usepackage{braket}
\usepackage[a4paper, textheight = 690pt, textwidth = 460pt, margin=0.75in]{geometry} 
\usepackage[dvipsnames]{xcolor}
\usepackage{bbm}
\usepackage{hyperref}
\hypersetup{
 colorlinks,
 citecolor=Red,
 linkcolor=Blue,
 urlcolor=Blue}

\newcommand{\cnub}{{C$\nu$B}}

\begin{document}

\singlespacing

{\hfill FERMILAB-PUB-18-374-T, NUHEP-TH/18-07}

\title{Multimessenger Astronomy and New Neutrino Physics}

\author{Kevin J. Kelly}
\affiliation{Northwestern University, Department of Physics \& Astronomy, 2145 Sheridan Road, Evanston, IL 60208, USA}
\author{Pedro A. N. Machado}
\affiliation{Theoretical Physics Department, Fermi National Accelerator Laboratory, P.O. Box 500, Batavia, IL 60510, USA}

\begin{abstract}
We discuss how to constrain new physics in the neutrino sector using multimessenger astronomical observations by the IceCube experiment. 
The information from time and direction coincidence with an identifiable source is used to improve experimental limits by constraining the mean free path of neutrinos from these sources. Over the coming years, IceCube is expected to detect neutrinos from a variety of neutrino-producing sources, and has already identified the Blazar TXS 0506+056 as a neutrino-producing source. We explore specific phenomenological models: additional neutrino interactions, neutrinophilic dark matter, and lepton-number-charged axion dark matter. For each new physics scenario, we interpret the observation of neutrinos from TXS 0506+056 as a constraint on the parameters of the new physics models. We also discuss mergers involving neutron stars and black holes, and how the detection of neutrinos coincident with these events could place bounds on the new physics models.
\end{abstract}


\maketitle

\setcounter{equation}{0}
\section{Introduction}
\label{sec:introduction}

The neutrino sector is the least known sector of the standard model -- the absolute value of their masses, the mass generation mechanism itself, and the Dirac or Majorana nature of neutrinos are a few of many open questions. Besides, long standing discrepancies in short baseline oscillation experiments~\cite{Aguilar:2001ty, Ko:2016owz, Aguilar-Arevalo:2018gpe, Alekseev:2018efk, Almazan:2018wln} are still to be understood, possibly pointing towards novel interactions secluded to the neutrino sector. In fact, current experimental constraints still allow relatively sizable new neutrino interactions, specially in scenarios where new physics is light and weakly coupled (see, e.g. Refs.~\cite{Gninenko:2009ks, Cherry:2014xra, Bai:2015ztj, Asaadi:2017bhx, Chu:2018gxk, Bertuzzo:2018itn}). As neutrinos interact very weakly with matter, exploring this sector is a challenging task, requiring non-trivial search strategies ranging from man made neutrino beams to astrophysical neutrino sources.

Since the discovery of ultra-high energy extraterrestrial neutrinos (UHE$\nu$) by IceCube~\cite{Aartsen:2013jdh}, a great deal of attention has been devoted to what can be learned about neutrino physics from these events. The underlying features that prompt such question are essentially the extreme conditions of UHE$\nu$  that are otherwise inaccessible at collider experiments: multi-PeV neutrinos traversing Gigaparsecs of cosmic neutrino and dark matter backgrounds to arrive at the Earth. The observation of the UHE$\nu$ spectrum provides an invaluable probe of physics beyond the standard model. For instance, the flavor composition of 
UHE$\nu$ can be used to probe non-standard neutrino interactions, mixing with sterile neutrinos and neutrino decays~\cite{Athar:2000yw, Keranen:2003xd, Blennow:2009rp, Mehta:2011qb, Hollander:2013im, Chatterjee:2013tza, Mena:2014sja, Palladino:2015zua, Arguelles:2015dca, Pagliaroli:2015rca, Bustamante:2015waa, Gonzalez-Garcia:2016gpq, Brdar:2016thq, Rasmussen:2017ert,Denton:2018aml}, as well as  CPT violation~\cite{Ando:2009ts, Barenboim:2003jm, Klop:2017dim,Ellis:2018ogq}; the shape of UHE$\nu$ energy spectrum can probe interactions between neutrinos and leptoquarks~\cite{Anchordoqui:2006wc, Barger:2013pla, Dutta:2015dka, Dey:2015eaa, Mileo:2016zeo, Chauhan:2017ndd, Dey:2017ede, Becirevic:2018uab};  and the arrival of extragalactic events is an indication that the universe is not opaque to neutrinos, constraining new interactions in the neutrino sector~\cite{Ioka:2014kca, Ng:2014pca, Cherry:2014xra, Shoemaker:2015qul, Araki:2015mya, DiFranzo:2015qea, Marfatia:2015hva, Davis:2015rza, Cherry:2016jol,Arguelles:2017atb,Chianese:2018ijk}.
In this paper, we will focus on the opacity of the universe to neutrinos, and how  multimessenger astrophysics can be used to extract additional information from IceCube UHE$\nu$. 

Our motivation relies on the following recent observations. First, the IceCube collaboration recently detected a high-energy neutrino event in time and directional correlation with a gamma ray flare from the blazar TXS 0506+056~\cite{IceCube:2018dnn,IceCube:2018cha}. Additionally, the IceCube collaboration, upon analyzing data from 2014-2015, has measured an excess of $13\pm 5$ events from the direction of TXS 0506+056 with $\sim$ TeV-PeV energy neutrinos. Measurements of TXS 0506+056 constrain its redshift to be $z = 0.3365 \pm 0.0010$~\cite{Paiano:2018qeq}, corresponding to a distance of $1.3$ Gpc, implying that the mean free path of neutrinos is larger than $1.3$ Gpc. Second, the LIGO collaboration has recently begun detecting gravitational waves (GW) from the merger of compact objects. The first such observation that detected GW coincident with electromagnetic (EM) followup from a wide range of frequencies~\citep{Evans:2017mmy,Goldstein:2017mmi,Savchenko:2017ffs,Pozanenko:2017jrn,Fermi-LAT:2017uvi,Coulter:2017wya,Chornock:2017sdf,Tanvir:2017pws,Nicholl:2017ahq,Soares-Santos:2017lru,Verrecchia:2017hck,Hu:2017tlb} was the event GW170817~\cite{TheLIGOScientific:2017qsa}. This has allowed the LIGO collaboration to estimate that there exists a non-zero rate of binary neutron star (NS-NS) mergers in the universe~\cite{TheLIGOScientific:2017qsa}of $1540^{+3200}_{-1220}$ Gpc$^{-3}$ yr$^{-1}$. While EM followup was detected for this, no neutrinos were detected coincident with GW170817 by a variety of neutrino detectors~\cite{ANTARES:2017bia}. Many have speculated about the possibility of high-energy neutrino emission from NS-NS mergers, and in the next generation of neutrino experiments, these types of events should be detectable~\cite{Dermer:2003zv, Waxman:1997ti, Murase:2013ffa, Moharana:2016xkz, Kimura:2017kan}. Neutrinos may also be emitted from Neutron Star-Black Hole (NS-BH) mergers, the rate of which in the universe is estimated to be between $0.5$ and $1000$ Gpc$^{-3}$ yr$^{-1}$~\cite{Abadie:2010cf}.
 
Our aim is to evaluate how much more one can learn about the neutrino sector using multimessenger astrophysics, that is, using the information from the coincidence  between a high energy neutrino signal and GW/EM observations. This coincidence bears valuable information, allowing to further assess the opacity of the universe to neutrinos as the direction, timing and distance of the neutrino source is identified. 
While the predicted neutrino fluxes from NS mergers and blazars vary based on a number of assumptions, making it hard to predict an event yield at a neutrino experiment, we may still use the possibility of detection to probe new physics in the neutrino sector. 
For example, if new neutrino interactions exist, there is the possibility that a neutrino emitted from an identified source can scatter off a neutrino from the Cosmic Neutrino Background (C$\nu$B), causing it to go undetected at Earth, or to have a neutrino signal delayed with respect to the optical one.

In this manuscript, we will focus on the detection of neutrinos coincident with the blazar TXS 0506+056 as a means of probing specific new physics scenarios. Additionally, we will discuss the future ability of IceCube Generation 2 (IceCube-gen2) and its sensitivity to NS mergers in probing these new physics scenarios. We analyze the existence of neutrino secret interactions ($\nu$SI), in which there exists a new massive particle coupling to neutrinos, the possibility of neutrinophilic dark matter, and the existence of a lepton-number-charged axion. We provide a recipe for setting limits on these scenarios given the detection of one neutrino event or more, and we discuss the possibility of discovering new physics in the neutrino sector in the absence of detecting neutrinos in particular regimes.

\section{Probing new physics with neutron star mergers and other multimessenger astrophysics}

For concreteness, we will focus on the neutrino signal associated with the blazar TXS 0506+056. Before going into details on how to constrain new interactions, we first discuss the general aspects of TXS 0506+056 that are relevant to our proposal. The blazar has been measured to be at a distance of $1.3$ Gpc from Earth~\cite{Paiano:2018qeq}, and a neutrino alert (IceCube-170922A) was detected coincident in time and location with a gamma-ray flare from the blazar. Subsequent analysis of historical IceCube data has uncovered an excess of $13\pm 5$ muon neutrino events in 2014 and 2015 from the direction of TXS 0506+056. The events in this time window are produced by neutrinos with energy $\sim$TeV-PeV~\cite{IceCube:2018cha}.

Interactions restricted to the neutrino sector are notoriously difficult to probe. However, neutrinos emitted in astrophysical environments pass through the cosmic neutrino background \cnub~with an average density of $n_\nu \sim$ 340 cm$^{-3}$, providing a promising environment to study neutrino--neutrino interactions. In the standard model, the mean free path of a PeV neutrino traversing the \cnub~is $\mathcal{O}(10^{11})$ Gpc. If a new interaction between neutrinos is present, the universe may eventually become opaque to high energy neutrinos, and thus the observation of neutrino events from an identifiable source can put strong constraints on $\nu-\nu$ interactions. Additionally, if sizable $\nu-\gamma$ or $\nu$-dark matter interactions exist, scattering off the Cosmic Microwave Background (CMB) or dark matter relic densities can cause a depletion of the detected neutrino events.

We assume that the observation of a neutrino event from an identifiable source implies that the mean free path of neutrinos $\lambda_\mathrm{MFP}$ with such an energy is greater than the progenitor distance $d$, which is measured precisely by EM/GW experiments. Specifically, we define the mean free path as
\begin{equation}
\lambda_\mathrm{MFP} = \frac{1}{n_X \sigma(\nu X \to Y)},
\end{equation}
where $n_X$ is the number density of particle $X$ (assumed to be  uniform) and $Y$ is the particle/particles produced by the interaction between $\nu$ and $X$. We perform this process for several new neutrino physics scenarios: ``secret neutrino interactions'' with a new mediator ($X = \nu/\bar{\nu}$), neutrinophilic dark matter ($X = \nu/\bar{\nu}$ or dark matter), and interactions between neutrinos and a relic density of axion dark matter ($X = a$).

Our simulation is as follows. For a given new physics hypothesis, we perform a $10^4$ pseudoexperiments attempting to replicate the IceCube measurement of events from TXS 0506+056. For a given pseudoexperiment, we draw a number (from a Gaussian distribution of $13 \pm 5$) of events from the published IceCube data between MJD of 56937.81 and 57096.21 (this range corresponds to the box method best-fit in Ref.~\cite{IceCube:2018cha} and contains 61 events)~\cite{ICData}. Each event's muon energy proxy $\hat{E}_\nu$ is published. We use the supplemental material (Figure S5) from Ref.~\cite{IceCube:2018cha} to estimate the most-likely value of true neutrino energy $E_\nu$ given $\hat{E}_\nu$, and arrive at the following relationship:
\begin{equation}
\left(\frac{E_\nu}{\mathrm{TeV}}\right) = 1.92 \left(\frac{\hat{E}_\nu}{\mathrm{TeV}}\right)^{1.14}.
\end{equation}
Following Figure S6 of the supplemental material from Ref.~\cite{IceCube:2018cha}, we approximate the uncertainty of $E_\nu$ to be an order-of-magnitude in width. We also assume a flat prior on $E_\nu$ over this range\footnote{A more thorough analysis of the results of TXS 0506+056 would include the event weights for each data point in Ref.~\cite{ICData}, and would also incorporate a more detailed simulation of the most likely value of $E_\nu$ given the measurement of $\hat{E}_\nu$, as well as the probability distribution of $E_\nu$. The aim of this manuscript is to provide motivation for such an analysis by showing its power in probing specific new physics models, and we encourage the experimental community to perform a more complete analysis}. The mean value of $\hat{E}_\nu$ in the time window we analyze is roughly $2.3$ TeV, corresponding to $E_\nu \sim 4.9$ TeV.

The observation of $13\pm 5$ neutrinos from TXS 0506+056 does not automatically imply that $\lambda_\mathrm{MFP}$ is greater than $1.3$ Gpc -- if the predicted number of neutrino events from the blazar flux were $\mathcal{O}(100)$ and neutrinos have a mean free path of $\mathcal{O}(500\ \mathrm{Mpc})$, the number observed would be reasonable. However, the measured neutrino luminosity in the time window used by IceCube is $1.2^{+0.6}_{-0.4} \times 10^{47}$ erg s$^{-1}$ compared to the gamma-ray luminosity of $0.28 \times 10^{47}$ erg s$^{-1}$~\cite{IceCube:2018dnn}. The two luminosities are expected to be similar, leading us to believe that $\lambda_\mathrm{MFP} > 1.3$ Gpc is a reasonable conclusion to draw.

In addition to neutrinos from blazars, it is expected in the coming years that IceCube (and its next-generation upgrade IceCube Generation 2) will be able to detect neutrinos associated with both GW and EM signatures from the mergers of neutron stars with either other neutron stars (NS-NS) or black holes (NS-BH)~\cite{Kimura:2017kan}. Over ten years of data collection, under optimistic astrophysical assumptions about the neutrino production in these environments, IceCube Generation 2 is expected to detect between roughly 20 and 50 neutrino events from the extended emission of the Short Gamma Ray Bursts (SGRBs) produced in the NS-NS or NS-BH merger. These events are identified to be coincident in direction and within a specified time interval of the GW/EM detection of the merger. IceCube will be most sensitive to events with neutrino energies $\sim 100$ TeV - $1$ PeV at distances of up to roughly 600 Mpc. For comparison with the search using TXS 0506+056, we display for each new physics scenario a dotted line corresponding to a constant mean free path of 600 Mpc for a 1 PeV neutrino. Uncertainties on a measured neutrino energy would smear the ability to set a limit on physics parameters, which is not accounted for here.

\subsection{Secret Neutrino Interactions and Neutrinophilic Dark Matter}
\label{sec:Secret}
Here we consider that neutrinos couple to a new massive particle $\phi$, where $\phi$ can be a scalar ($\phi=S$), pseudoscalar ($P$), vector ($V$), or axial-vector ($A$) particle. Assuming the coupling to this new particle to be  $g_X$, the cross sections for neutrino scattering via this new particle are
\begin{align}
\sigma(\nu\bar{\nu}\to S \to \nu\bar{\nu}) &\simeq \frac{g^4 m_\phi^4 (1-\epsilon)}{64\pi s(s-m_\phi^2)^2} \frac{s^2((2-\epsilon)\epsilon + 4) - 4sm_\phi^2 + 8m_\phi^4}{(2(s+m_\phi^2) - \epsilon s)(\epsilon s + 2m_\phi^2)} + \frac{g^4 m_\phi^4}{16\pi s^2(s-m_\phi^2)} \tanh^{-1}{\left(\frac{s(1-\epsilon)}{s+2m_\phi^2}\right)}, \\
\sigma(\nu\bar{\nu}\to P \to \nu\bar{\nu}) &\simeq \sigma(\nu\bar{\nu}\to S \to \nu\bar{\nu}).
\end{align}
\begin{align}\label{eq:sigma-V}
\sigma(\nu\bar{\nu} \to V \to \nu\bar{\nu}) &\simeq \frac{g^4 m_\phi^2 (s+m_\phi^2)}{4\pi s^2 (s-m_\phi^2)} \log{\left[\frac{(2-\epsilon)s + 2m_\phi^2}{\epsilon s + 2m_\phi^2}\right]} + \frac{g^4 (\epsilon -1) P^{(8)}(s, m_\phi, \epsilon)}{192\pi s(s-m_\phi^2)^2 (\epsilon s + 2m_\phi^2)((\epsilon-2)s - 2m_\phi^2)},\\
\sigma(\nu\bar{\nu} \to A \to \nu\bar{\nu}) &\simeq \sigma(\nu\bar{\nu} \to V \to \nu\bar{\nu}), \\
P^{(8)}(s, m_\phi, \epsilon) &\simeq s^4(48 + \epsilon(2-\epsilon)(\epsilon(\epsilon-2)-20)) + 16s^3 m_\phi^2 (\epsilon(\epsilon-2) - 5) \nonumber \\
&\quad- 32s^2 m_\phi^4 (\epsilon(\epsilon-2) + 7) + 96s m_\phi^6 + 192 m_\phi^8.
\end{align}
In the $m_\nu \to 0$ limit, the distinctions between scalar/pseudoscalar and vector/axial vector cross sections vanish. The parameter $\epsilon$ corresponds to a minimum center-of-mass-frame scattering angle to regularize this cross section, namely
\begin{equation}
  \sigma_{\rm total} = \int_{-1+\epsilon}^{1-\epsilon} d(\cos\theta)\frac{d\sigma}{d\cos\theta}.
\end{equation}
Without non-zero $\epsilon$, the cross section does not goes to zero in the $s\to\infty$ limit. We use $\epsilon=0.05$, however the numerical results do not depend strongly on this value. There also exists a $t$-channel scattering amplitude in each of these cases, however the cross section due to this is negligible compared to those shown.

\begin{figure}[t]
\centering
\hspace{-0.5cm}
\includegraphics[width=0.52\linewidth]{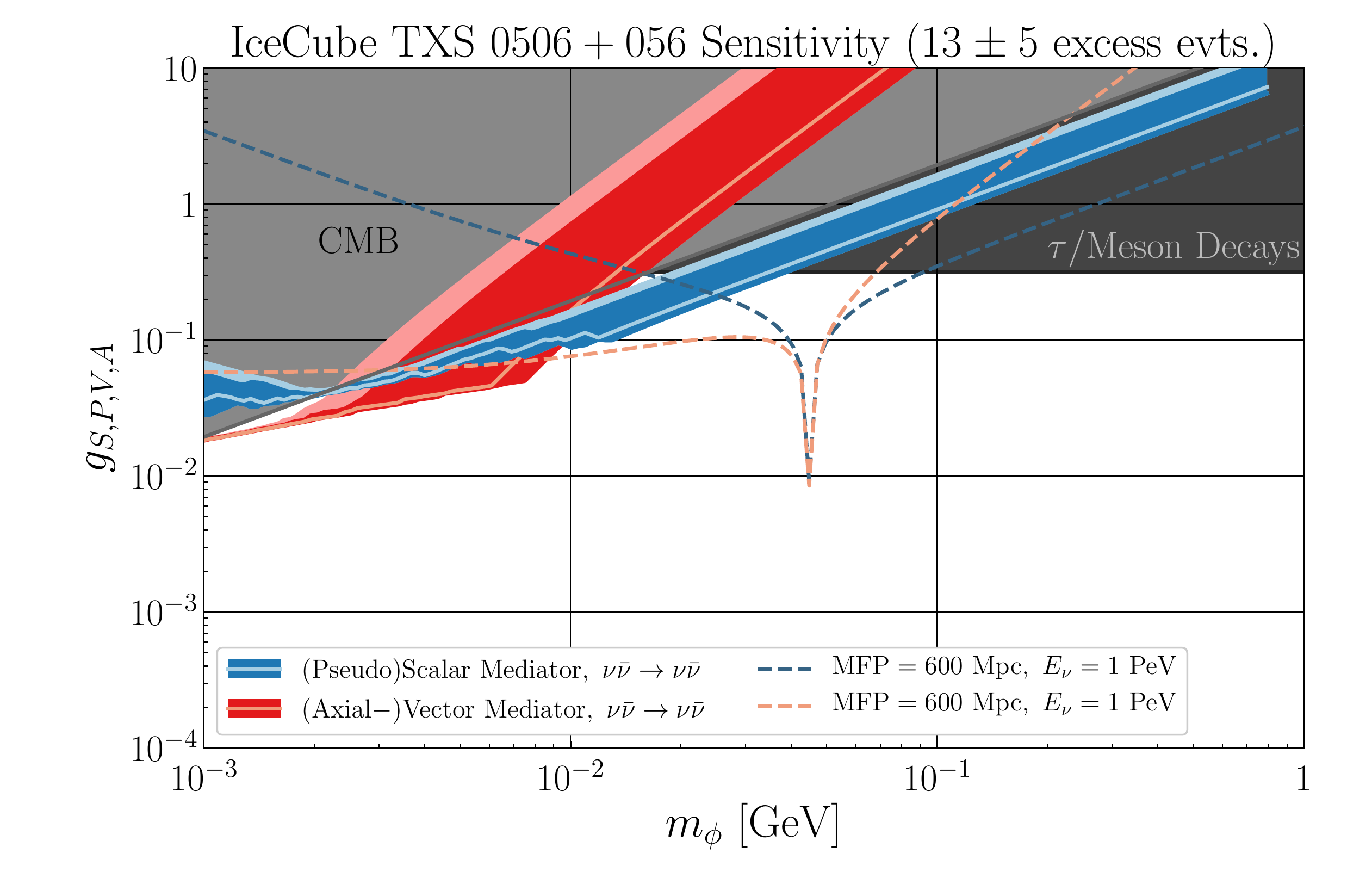}\hspace{-0.5cm}
\includegraphics[width=0.52\linewidth]{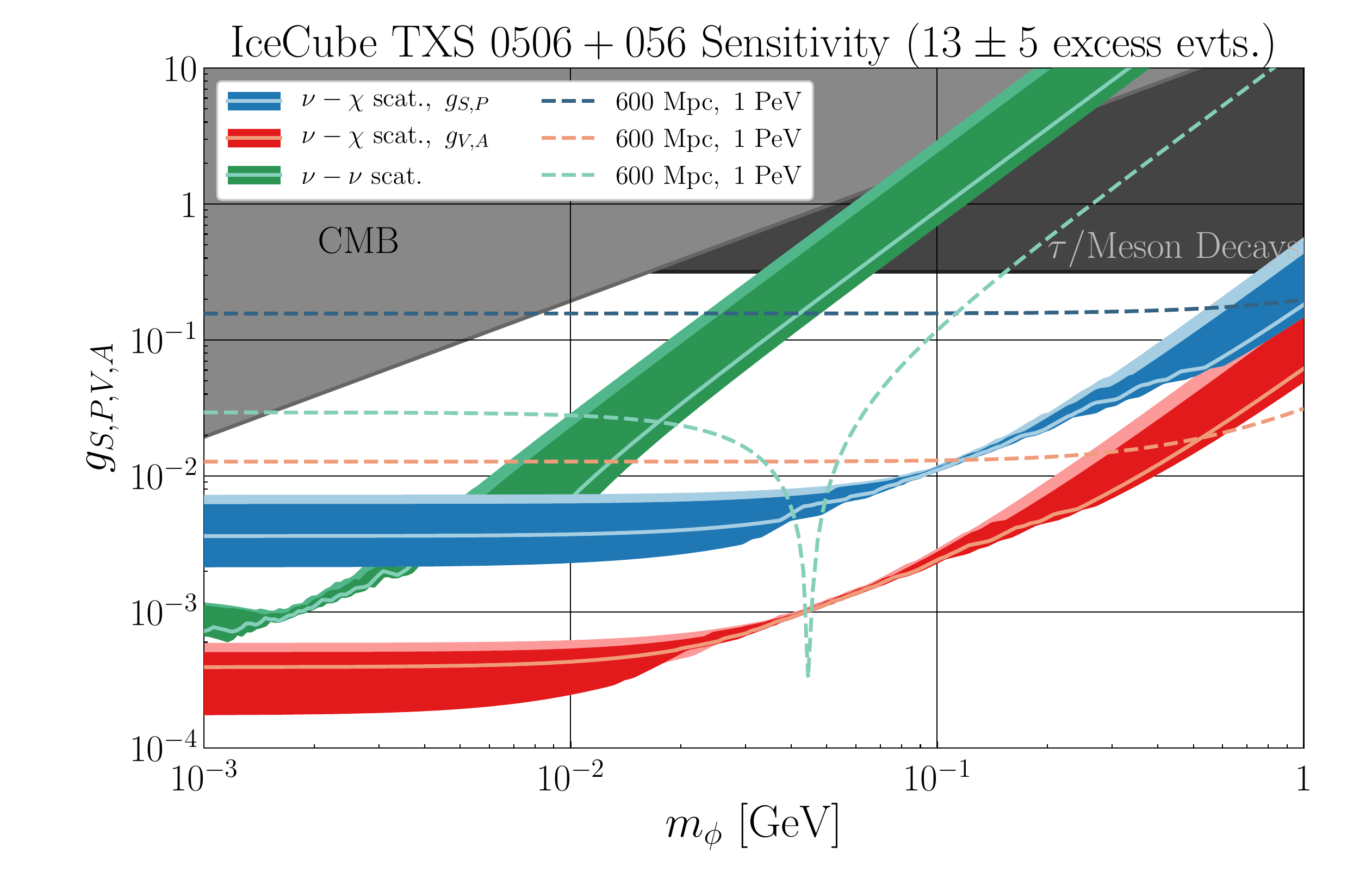}
\caption{Expected limits on secret neutrino interactions (left) or neutrinophilic dark matter (right) from $13\pm 5$ excess signal events from the blazar TXS 0506+056. We show expected limits assuming the new mediator $\phi$ is a scalar/pseudoscalar (blue) and a vector/axial-vector (red) in each panel. We also show existing bounds on the neutrino/$\phi$ coupling $g_{S,P,V,A}$ from cosmological probes\cite{Cyr-Racine:2013jua,Archidiacono:2013dua} and $\tau$/meson decays~\cite{Lessa:2007up} in gray. In the right figure, we show limits from neutrino scattering off relic $\chi$ dark matter in red and blue, and also resonant scattering $\nu \bar{\nu} \to \phi \to \chi\bar{\chi}$ in green. Here, we assume the $\chi$/$\phi$ coupling $g_D = 1$ and that $m_\chi = 5$ keV.}
\label{fig:SPVASensitivity}
\end{figure}

Fig.~\ref{fig:SPVASensitivity} displays our estimate of the IceCube limits to secret neutrino interactions from the $13\pm 5$ excess events coincident in location with TXS 0506+056. Limits are displayed as a function of $m_\phi$ and $g_{S,P,V,A}$, the scalar, pseudoscalar, vector, or axial-vector coupling between $\phi$ and neutrinos. The cross-sections for scalar/pseudoscalar and vector/axial-vector interactions are identical in the limit $m_\nu \to 0$, so only two limit regions are shown. Scalar/pseudo-scalar limits are shown in blue, and vector/axial-vector limits are shown in red. The line in the center of each band corresponds to the median expected limit, and the dark (light) surrounding region corresponds to the $1\sigma$ (95\% CL) limit from performing $10^4$ pseudoexperiments, as discussed above. We include constraints on these couplings from the CMB~\cite{Cyr-Racine:2013jua,Archidiacono:2013dua} and the weakest (in terms of flavor dependence) bound from meson and charged-lepton decays~\cite{Lessa:2007up}.
For simplicity, we assume the \cnub~neutrinos have a mass of $1$ eV and are at rest in the universe. 
The strictest bound on $g_{S,P,V,A}$ is set for
\begin{equation}
m_\phi \simeq \sqrt{2 E_\nu m_\nu} \simeq 14\ \mathrm{MeV}\ \left(\frac{E_\nu}{100\ \mathrm{TeV}}\right)^{1/2} \left(\frac{m_\nu}{1\ \mathrm{eV}}\right)^{1/2}.
\end{equation}
We note here that bounds placed by IceCube can also apply to neutrinos scattering and producing a lepton-number-charged scalar, as the \cnub~consists of both neutrinos and antineutrinos, so the $s$-channel process $\nu\nu\to\phi\to\nu\nu$ can occur~\cite{Berryman:2018ogk}.

If a new particle that couples to neutrinos can serve as a dark matter mediator, we can further probe the $g_{S,P,V,A}$ vs. $m_\phi$ parameter space (right panel of Fig.~\ref{fig:SPVASensitivity}). Assuming $\phi$ couples to a fermionic DM $\chi$ with a coupling  $g_D = 1$, there are two contributions to the neutrino mean free path: $\nu\bar{\nu}$ annihilation into $\chi\bar{\chi}$ via an $s$-channel\footnote{This cross section is largely insensitive to $m_\chi$ as long as the $\chi\bar{\chi}$ final state is kinematically accessible, or roughly $m_\chi \lesssim \sqrt{E_\nu m_\nu/2}$. If $E_\nu$ is 100 TeV and $m_\nu$ is 1 eV, this requires $m_\chi \lesssim 7$ MeV.} mediator, or $t$-channel scattering off relic $\chi$ dark matter. The right panel of Fig.~\ref{fig:SPVASensitivity} also displays expected limits on $g_{S,P,V,A}$ under these assumptions with $m_\chi = 5$ keV as an example. In contrast to the secret interactions case, when the target is a vector-like keV dark matter particle, the Lorentz structure of the coupling  leads to a different cross section due to the non-zero mass of the dark matter. The difference between scalar, pseudoscalar, vector, and axial-vector $\phi$ is negligible for the $\nu\nu$ scattering (green) in this case.

Finally, for both secret interactions and neutrinophilic DM cases, we show a dashed line of constant mean free path of 600~Mpc for a 1 PeV neutrino, the rough distance and energy expected in neutrino events from NS mergers. Although this is not an expected constraint (as one would need to consider energy uncertainties, etc.), the lines serve to give an idea of the impact of a higher-energy neutrino event on the sensitivity to a new physics model, even despite the shorter mean free path compared with TXS 0506+056. The future detection of neutrinos associated with NS merger events will allow us to probe heavier mediator masses $m_\phi$ than with TXS 0506+056 alone. Notice that the scalar and pseudoscalar lines behave differently from the vector and axial-vector for the secret neutrino interaction (Fig.~\ref{fig:SPVASensitivity} left panel) in the limit $m_\phi\to 0$. This is due to a flattening of the V/A cross section as $m_\phi$ goes to zero (second term in Eq.~(\ref{eq:sigma-V})), compared to a behavior $\propto g^4 m_\phi^2$ for the S/P case.

\subsection{Axion-Neutrino Couplings}
\label{sec:Axion}

Axions provide a compelling solution  to the strong $CP$ problem via the Peccei-Quinn mechanism, which also predicts a bosonic dark matter candidate with sub-eV masses~\cite{Peccei:1977hh}. The field that breaks the Peccei-Quinn symmetry, if it carries lepton number,  could also be responsible for the Majorana mass of right-handed neutrinos $N$, having thus a larger coupling to these hypothetical fermions~\cite{Mohapatra:1982tc}.
  In that case, neutrinos would couple to the axion via $\nu-N$ mixing. The corresponding Lagrangian is
\begin{equation}
\mathcal{L}_\nu = y_\nu \overline{L} \widetilde{H} N + y_N \phi \overline{N^c} N + \mathrm{h.c.},
\end{equation}
where $\phi$ is the Peccei-Quinn scalar (whose phase is the axion $a$).
To be model independent, we consider separate couplings between axions and the active neutrinos ($\mathcal{L} \supset g_{a\nu\nu} a \overline{\nu} \gamma^5 \nu$) and between axions and one active neutrino and one heavy neutrino ($\mathcal{L} \supset g_{a\nu N} a \overline{\nu} \gamma^5 N$), however, in specific realizations of the model,  these couplings are related by the neutrino masses and mixings.

Assuming only the coupling $g_{a\nu\nu}$ exists, astrophysical neutrinos may scatter off relic axions with the cross-section 
\begin{align}
\sigma(a + \nu \to a + \nu) = \frac{g_{a\nu\nu}^4 s}{64\pi(s-m_\nu^2)^2} \left[ 1 + \frac{17m_\nu^2 - 4m_a^2}{s} + \frac{14m_\nu^4 - 12m_\nu^2 m_a^2 + 2m_a^4}{s^2}\right].
\end{align}
Assuming that the axions comprise the relic dark matter density in the universe ($\rho_\mathrm{DM} = 0.3$ GeV/cm$^3$) and that the axion velocity is $v_a \sim 10^{-3}$, constraining the mean free path of neutrinos to be above a distance $d$ allows us to place a bound on $g_{a\nu\nu}$ of
\begin{equation}\label{eq:axion-constraint}
g_{a\nu\nu} \lesssim 1.83 \times 10^{-6} \left(\frac{E_\nu}{10\ \mathrm{TeV}}\right)^{1/4} \left(\frac{m_a}{10^{-6}\ \mathrm{eV}}\right)^{1/2}\left(\frac{d}{1\ \mathrm{Gpc}}\right)^{-1/4}.
\end{equation}
We can trace back each term in this inequality by looking at the cross section and mean free path formula. The first is simply the cross section that goes as $g_{a\nu\nu}^4E_\nu$. The second is a combination of the cross section behavior ($\sigma\propto g_{a\nu\nu}^4/m_a$) and axion number density, while the last is simply due to the mean free path definition, as $\lambda_{\rm MFP}\propto 1/g_{a\nu\nu}^4$.
This coupling is also constrained by particle emission in double-beta decay experiments~\cite{Gando:2012pj}, giving $g_{a\nu_e \nu_e} \lesssim 10^{-5}$, and by Planck measurements including free streaming and $N_\mathrm{eff}$~\cite{Friedland:2007vv}, which give $g_{a\nu\nu} \lesssim 10^{-13}(1~{\rm eV}/m_a)^2$. Notice that our constraint (\ref{eq:axion-constraint}) is complementary to the $N_\mathrm{eff}$ one, with different dependence on the mass $m_a$.

\begin{figure}[!tbp]
\centering
\includegraphics[width=0.7\linewidth]{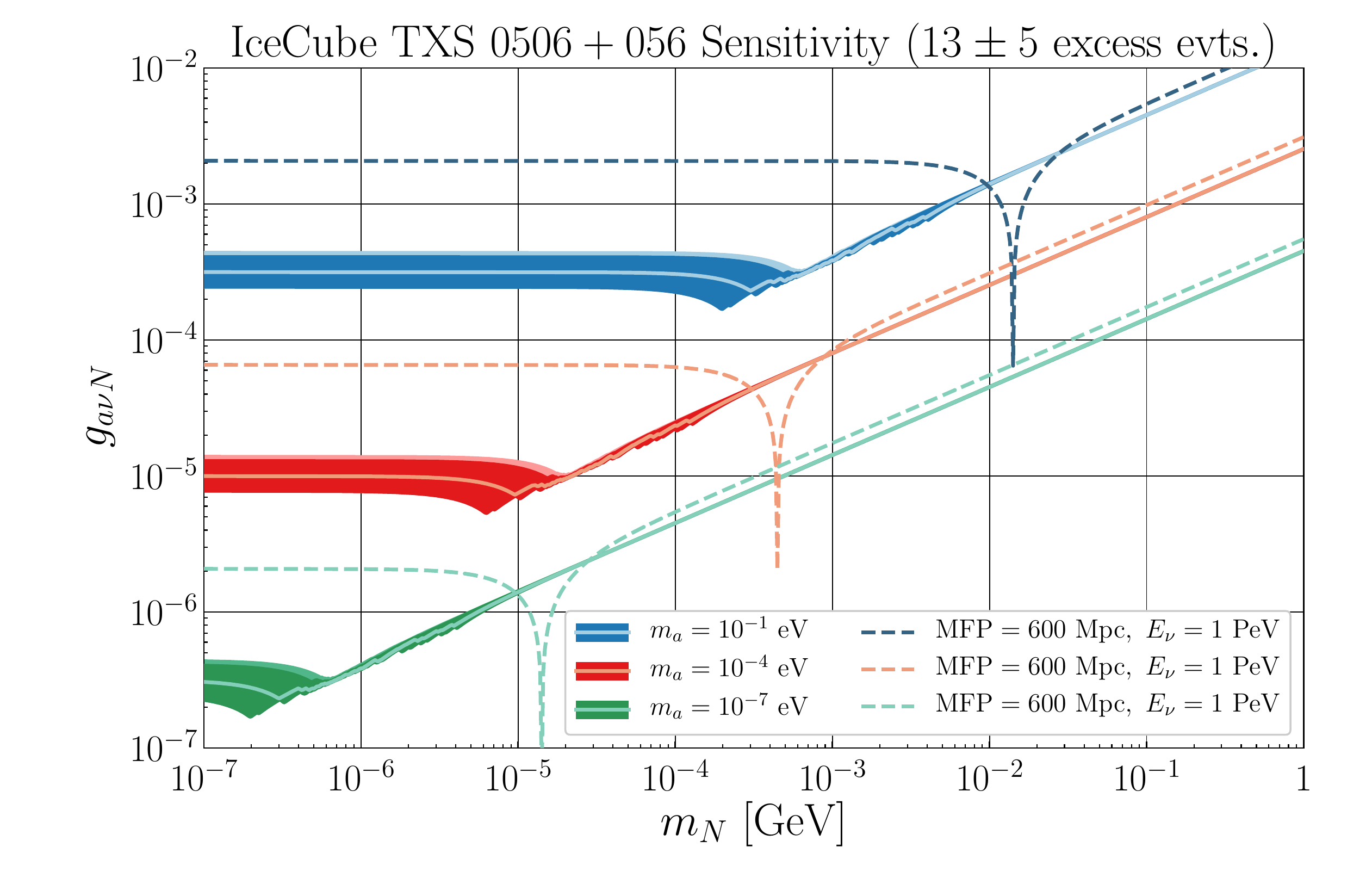}
\caption{Expected limits from $13\pm 5$ excess signal events from the blazar TXS 0506+056 assuming relic axions couple to active neutrinos $\nu$ and a heavy neutrino $N$ (with mass $m_N$) via a coupling $g_{a\nu N}$. We show limits for three choices of axion masses: $10^{-1}$ eV (blue), $10^{-4}$ eV (red), and $10^{-7}$ eV (green). The line at the center of each region corresponds to median expected limits, dark regions correspond to $\pm 1\sigma$ expectation, and light regions $95\%$ CL. For comparison, we show dashed lines that correspond to a constant mean free path of $600$ Mpc for $1$ PeV neutrinos, roughly the distance and energy of NS merger neutrinos.}
\label{fig:AxionSensitivity}
\end{figure}

With the coupling $g_{a\nu N}$, this process will include an $s$-channel heavy neutrino $N$ with the cross section
\begin{align}
\sigma(a + \nu \to N \to a + \nu) = \frac{g_{a\nu N}^4 s}{64\pi(s-m_N^2)^2} &\left[1 + \frac{m_N^2 + 8m_N m_\nu + 8 m_\nu^2 - 4m_a^2}{s}\right.  \label{eq:anuN}\\
 &\left. + \frac{4m_N^2 m_\nu^2 + 8 m_N m_\nu^3 + 2 m_\nu^4 - 8 m_N m_\nu m_a^2 - 4m_a^2 m_\nu^2 + 2m_a^4}{s^2} \right].\nonumber
\end{align}
Depending on the axion mass, one can set limits in the $m_N - g_{a\nu N}$ plane, which we display in Fig.~\ref{fig:AxionSensitivity}.
Limits are better for lighter axion masses (where the preferred axion mass is near $10^{-6}$~eV~\cite{Mohapatra:1982tc}). Heavy neutrino masses on the order of keV-MeV are accessible in this scenario.

As with the secret neutrino interactions and neutrinophilic dark matter, we show a line of constant mean free path of $600$ Mpc for a neutrino energy of $1$ PeV, associated with the distance and energy of neutrinos from NS mergers that could be detected by IceCube Generation 2. We see that, except for in the region $s \simeq m_N^2$, the limit set by TXS 0506+056 is stronger than this example: this is because the cross section in Eq.~(\ref{eq:anuN}) is larger for the lower-energy neutrinos from TXS 0506+056.

We can understand the flat (linear) behavior of the curves in Fig.~\ref{fig:AxionSensitivity} for small (large) $m_N$ as follows. As $m_N \to 0$, the cross section in Eq.(\ref{eq:anuN}) goes to $g_{a\nu N}^4/(128\pi m_a E_\nu)$, eliminating any $m_N$ dependence. The spread in the limits from TXS 0506+056 for low $m_N$ then comes from the spread in energies of the detected neutrinos. Once $m_N^2 >> s$, the cross section becomes $g_{a\nu N}^4/(64\pi m_N^2)$, now independent of neutrino energy. There is still $m_a$ dependence on the limits as $n_a$, the axion number density, is inversely proportional to $m_a$ -- explaining why bounds are stronger for low $m_a$ than high.

\subsection{Absence of Neutrinos above some Distance}
In general, it is easier to interpret the detection of a neutrino (coincident with an identified source) as a lower limit on the neutrino mean free path than it is to observe no neutrinos from some source and argue that all neutrinos have been absorbed en route to Earth. This is due in large part to the uncertainties on the neutrino production by astrophysical sources, such as neutron star mergers~\cite{Kimura:2017kan}.

As discussed in Ref.~\cite{Kimura:2017kan}, several tens of neutrinos from neutron star merger events are predicted to be detected after ten years of IceCube Generation 2. If no neutrinos are detected in this timespan, either (a) models predicting neutrino production from SGRBs must be adjusted or (b) new physics, such as neutrino absorption discussed here, is at play. However, suppose neutrinos are detected associated with several NS merger events below a given distance, but not in several events with more distant progenitors. Then, (b) is the most likely explanation. If the detected neutrinos have energy $\sim 1$ PeV and the distance above which no events are detected is $\sim 600$ Mpc, then the dashed lines in Figs.~\ref{fig:SPVASensitivity} and \ref{fig:AxionSensitivity} depict the preferred region of parameter space for this explanation.

\section{Conclusions}

We have discussed how new physics in the neutrino sector can be probed with multimessenger astronomical observations in IceCube and its upgrade, Generation 2. By identifying the neutrino-producing source with time and direction coincidence with gravitational wave and/or electromagnetic observations, we may improve on constraints on the opacity of the universe to neutrinos. To exemplify this method, we have performed an analysis of the neutrino events detected coincident with the blazar TXS 0506+056 and interpreted their detection in several simplified models: neutrino secret interactions, neutrinophilic dark matter, and axion-neutrino couplings. For the first two models, a more thorough analysis than ours would improve the bounds from the CMB and $\tau$/meson decays. For the axion case, we derive complementary constraints to cosmological bounds. IceCube could also place strong bounds for sub-micro-eV axions, if the right-handed neutrinos are in the keV-MeV scale.

Briefly, we offer some remarks on new neutrino physics scenarios in which utilizing the large distances traveled by ultra-high energy extraterrestrial neutrinos is not advantageous compared with other methods of neutrino detection. Famously, the observation of neutrinos from Supernova 1987A has been used to place constraints on new physics models, such as the existence of a neutrino magnetic moment~\cite{Barbieri:1988nh}, neutrinos with nonzero electric charge~\cite{Barbiellini:1987zz}, or new particle interactions~\cite{Raffelt:1987yt}. Unlike the scenarios studied in this work, these specific models are more easily probed by low-energy ($E_\nu \sim$ several MeV) neutrinos. As an explicit example, if one were to attempt to constrain the lifetime of neutrino decay, the Lorentz factor $\gamma = E_\nu/m_\nu$ makes probes with ultra-high energy extraterrestrial neutrinos feeble in comparison to those from distant supernovae.

Having identified specific new neutrino physics models that are better probed by high-energy neutrinos, we emphasize the importance of multimessenger astronomy in searching for this new physics. Of critical importance here is the ability to pinpoint the sources of neutrinos, both in terms of time and direction. Then, and only then, can one be confident that the mean free path of travel is greater than the distance of the observation of gravitational waves and/or electromagnetic signature detection. Over the next decades, the joint effort of these probes will be able to explore a wide variety of new physics models in the neutrino sector. 

\section*{Acknowledgements}
We thank Andr\'{e} de Gouv\^{e}a for useful discussions. 
Fermilab is operated by Fermi Research Alliance, LLC, under Contract 
No. DE-AC02-07CH11359 with the US Department of Energy.  PM acknowledges support from the EU 
grants H2020-MSCA-ITN-2015/674896-Elusives and H2020-MSCA-2015-690575-InvisiblesPlus. KJK thanks the Fermilab Neutrino Physics Center for support during work on this manuscript. The work of KJK is supported in part by Department of Energy grant \#de-sc0010143.


\bibliographystyle{apsrev-title}
\bibliography{BNSBib}{}

\end{document}